\documentclass{Interspeech2024}

\interspeechcameraready 
\usepackage[cmyk]{xcolor}
\usepackage{multirow}
\usepackage{float}
\usepackage{booktabs}
\usepackage{cite}
\usepackage{hyperref}
\usepackage{verbatim}
\hypersetup{
  colorlinks=true,    
  linkcolor=black,     
  citecolor=black,    
  urlcolor=black 
}

\newcommand{\astfootnote}[1]{%
  \begingroup
  \renewcommand\thefootnote{\fnsymbol{footnote}}%
  \footnotetext[1]{#1}%
  \endgroup
}
\usepackage[misc]{ifsym}

\title{FakeSound: Deepfake General Audio Detection}

\name{Zeyu}{Xie}
\name{Baihan}{Li}
\name{Xuenan}{Xu}
\name{Zheng}{Liang}
\name{Kai}{Yu$^{\ast}$}
\name{Mengyue}{Wu$^{\ast}$}

\address{
   MoE Key Lab of Artificial Intelligence
   X-LANCE Lab, \\
   Department of Computer Science and Engineering
   AI Institute, \\
   Shanghai Jiao Tong University, Shanghai, China}
\email{\{zeyu\_xie, lbh0612, wsntxxn, liangzhenglz, kai.yu, mengyuewu\}@sjtu.edu.cn}


\begin{document}

\maketitle

\begin{abstract}
    
With the advancement of audio generation, generative models can produce highly realistic audios.
However, the proliferation of deepfake general audio can pose negative consequences. 
Therefore, we propose a new task, deepfake general audio detection, which aims to identify whether audio content is manipulated and to locate deepfake regions. 
Leveraging an automated manipulation pipeline, a dataset named FakeSound for deepfake general audio detection is proposed, and samples can be viewed on website \href{https://FakeSoundData.github.io}{\color{cyan}{\textit{https://FakeSoundData.github.io}}}. 
The average binary accuracy of humans on all test sets is consistently below 0.6, which indicates the difficulty humans face in discerning deepfake audio and affirms the efficacy of the FakeSound dataset.
A deepfake detection model utilizing a general audio pre-trained model is proposed as a benchmark system.
Experimental results demonstrate that the performance of the proposed model surpasses the state-of-the-art in deepfake speech detection and human testers.
\end{abstract}

\section{Introduction}
\astfootnote{Mengyue Wu and Kai Yu are the corresponding authors.}
Recently, generative artificial intelligence has witnessed rapid development, with models capable of generating highly realistic images and speech. 
However, there is a potential threat if these technologies are misused by malicious actors to harm society, leading to significant societal risks.
The field of computer vision has recognized this issue and proposed DeepFake Detection Challenge (DFDC)\cite{dolhansky2020deepfake} to identify whether a particular video segment contains deepfake frames manipulated by models.
Similarly, speech deepfake detection has emerged as a new research topic, including challenges such as the Automatic Speaker Verification Spoofing and Countermeasures Challenge (ASVspoof 2021)\cite{yamagishi2021asvspoof} and Audio Deepfake Detection Challenge (ADD 2022, ADD 2023)~\cite{yi2022add, yi2023add}, which have played a crucial role in promoting research in deepfake speech detection field.

Nevertheless, general audio deepfake detection has received little attention.
General audio encompasses any audio content including environmental sound, speech, etc., featuring a wider range of categories, more diverse content, and typically diverse audio quality compared to standard speech audio. 
Particularly, general audio typically lacks linguistic, rhythmic, and tonal information exhibited in speech audio, making detection more challenging than deepfake speech detection.

With advancements in audio generation models, general audio that is nearly indistinguishable from human-generated content can be synthesized~\cite{ghosal2023text,huang2023make,kreuk2022audiogen,liu2023audioldm2,yang2023diffsound,vyas2023audiobox,yang2023uniaudio}. 
These deepfake general audio files may be misused, leading to societal problems such as the dissemination of fake news, audio-based scams, falsification of legal evidence, enhanced deception in fake videos, and decreased credibility of digital information.
Therefore, we propose deepfake general audio detection to encourage researchers to focus on and delve deeper into deepfake audio detection technology.

\begin{figure}[t]
  \centering
  \centerline{\includegraphics[width=0.93\linewidth]{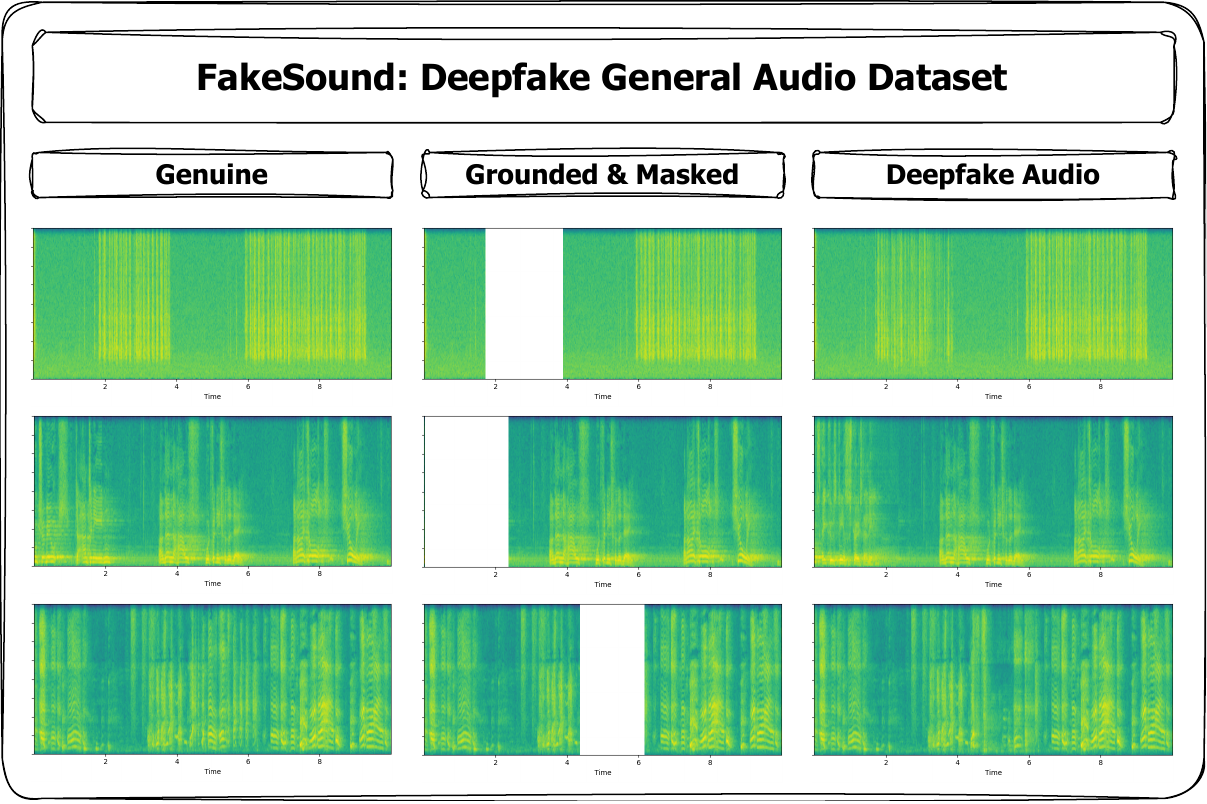}}
  \caption{
  Samples of FakeSound, synthesized by the manipulation pipeline. 
  A grounding model locates and masks key regions in the genuine audio, followed by regeneration and replacement by the generation model.
  }
  \label{fig:intro_example}
\end{figure}

Deepfake general audio detection aims to identify whether any audio content is manipulated and to locate fake regions. 
There are several types of fake audio: 1) the entire clip is regenerated; 2) some segments are spliced with another; 3) some segments filled in by a generative model through inpainting. 
The last ``half-truth" type is the most challenging to detect because it contains both genuine and generated segments.
Even humans find it difficult to discern when the inpainting model performs well.
The average accuracy of humans in identifying deepfake audio is below $0.6$, as illustrated by subjective evaluation in Table~\ref{tab:result}.
Thus, we focus on the most difficult fakeaudio, half-truth deepfake general audios, wherein certain segments are generated by inpainting models.

\begin{table*}[t]
    \renewcommand{\arraystretch}{1.2}
    \centering
   
    \caption{ 
        The metadata of FakeSound dataset. 
        \textbf{Manipulated Segment Limit} indicates the limit on the duration for regenerating key regions.
        \textbf{Inpainting Model} refers to the generative model used for inpainting.
    }
    \begin{tabular}{c|cccc}
    \toprule
    Dataset & Train & Test-Easy & Test-Hard & Test-Zeroshot   \\
   \midrule
    Number of Simulated Instances       &3,166       &92         &270      & 270               \\
    \midrule
    Manipulated Segment Limit  & 1-4 second      & 1-4 second      &Unlimited  &Unlimited         \\
   \midrule
    \multirow{2}{*}{Inpainting Model}  &AudioLDM2&AudioLDM2         &AudioLDM2 &\multirow{2}{*}{AudioLDM1}      \\
    &+AudioSR &+AudioSR & +AudioSR &\\
    \bottomrule
    \end{tabular}
    \label{tab:dataset}
\end{table*}

However, there is currently no dataset available specifically for the task of detecting deepfake general audio.
A preceding speech deepfake dataset employed a text-to-speech model to generate and subsequently replace several words within audio clips, resulting in the curation of the Half-truth Speech dataset~\cite{yi2021half}.
Following previous adoptions in speech, we design an automated manipulation pipeline specifically for general audio. 
This pipeline utilizes high-performing grounding, regeneration, and super-resolution models to efficiently generate deepfake general audios.
A deepfake general audio dataset, \textbf{FakeSound}\footnote[1]{
The FakeSound dataset, along with the training and evaluation code, is available at \href{https://github.com/FakeSoundData/FakeSound}{\color{cyan}{\textit{https://github.com/FakeSoundData/FakeSound}}} while the new ``fake" types and the generative models used will continue to be updated.}, is proposed for training and comprehensive evaluation of the deepfake general audio detection model.
We also propose a deepfake detection model as a benchmark system.
Experimental results demonstrate that proposed model outperforms the state-of-the-art model (SOTA) in the speech deepfake detection competition and human evaluators. 



\section{FakeSound: Deepfake General Audio Dataset}
\label{sec:dataset}
A deepfake audio benchmark dataset requires inclusion of numerous fake scenarios such as various sound types, different generation systems, etc. and preferably with exact annotations. 
To largely avoid human involvement, we propose a manipulation pipeline to automate deepfake audio generation, as illustrated on the left side of Figure~\ref{fig:model}.

\subsection{Ground \& Mask}
To construct a fakesound dataset, we need sound events with precise timestamps. These single-sourced segments are considered key segments.
As these key segments contain the most crucial information of the audio, any alteration to them would have the most significant impact on the content of the audio. 
Therefore, we first employ an audio-text grounding model~\cite{xu2024towards} to locate the key segments of the audio, as it is less sensitive to threshold compared with sound event detection models. 
The grounding model detects the region that is highly correlated with the given text, while simultaneously filtering out audios.
After obtaining the key segments, we randomly select one of them and mask it with zeros.
For example, for a clip corresponding to caption ``someone is typing on a keyboard", the grounding model locates $N$ segments containing ``keyboard" sounds, among which one random segment is masked.

\subsection{Regenerate \& Replace}
After masking the key regions of the original audio, the generation model regenerates them. 
Open-source models such as AudioLDM1/2~\cite{liu2023audioldm1, liu2023audioldm2} provide inpainting of the masked portions based on input text and the remaining audio information.

To further enhance the realism of the regenerated segments and ensure their quality, AudioSR~\cite{liu2023audiosr} is used for upsampling. 
Finally, the regenerated segments are concatenated with the original audio to cover the masked key segments.

\subsection{Dataset Metadata}
Our dataset FakeSound employs AudioCaps~\cite{kim2019audiocaps}, a widely utilized dataset in text-to-audio generation task, for deepfake general audio manipulation.
The first caption corresponding to audio clip is used as the text prompt for grounding and inpainting. 
AudioLDM2 and AudioSR are utilized as the inpainting model for simulating training set.
To ensure the quality of deepfake audio, the manipulated regions are limited to $1$ to $4$ seconds. 
Longer segments may degrade the quality of the generated audio, while shorter segments may not introduce significant changes to the original audio, which is disadvantageous for model learning.
If no segments within this range are detected by the grounding model, they will be filtered out from the training set.
To comprehensively evaluate the model performance, we manipulated $3$ test sets:
\begin{enumerate}
    \item \textit{Test-Easy} dataset is consistent with the settings of the training set, measuring the models' deepfake detection capabilities under the same data distribution;
    \item \textit{Test-Hard} dataset relaxes the constraint on the generated region being between $1$ to $4$ seconds. It contains audio samples of arbitrary manipulated duration, with arbitrary event length changes, and varying levels of generated quality. This dataset is used to assess the model's deepfake detection capabilities in complex scenarios. We expect this to be a more difficult setting for both model and human evaluation.
    \item \textit{Test-zeroshot} goes a step further than Test-Hard by utilizing a distinct inpainting model AudioLDM1, which has not been used for simulating training data. 
\end{enumerate}



\begin{figure*}[htbp]
  
  \centering
  \centerline{\includegraphics[width=1\textwidth]{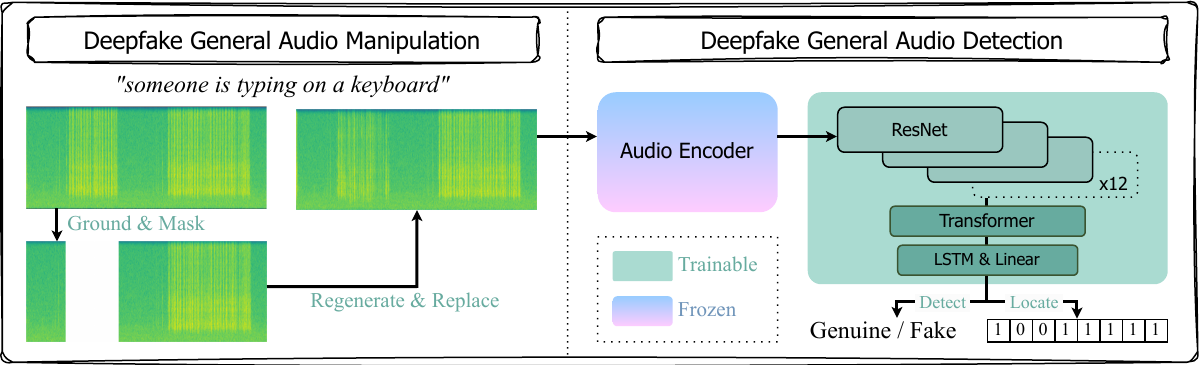}}
  
  \caption{
  \textbf{Left}: Manipulation pipeline. 
  A grounding model locates and masks key regions on genuine audio based on caption information. 
  The generation model regenerates these key regions, replacing them to produce convincing realistic 
  deepfake general audio.
  \textbf{Right}: Diagram of proposed model, which conducts deepfake detection on input general audio\textemdash identifies whether the audio is genuine or deepfake, and locates the deepfake regions.
  \label{fig:model}
}
\end{figure*}

\section{Evaluation Metric}
Deepfake general audio detection requires the model to (1) identify whether the audio is genuine or deepfake, and (2) locate the deepfake regions.
Following the setup of ADD2023~\cite{yi2023add}, a detection $Score$ is introduced, which is a composite metric calculated as the weighted sum of identification accuracy $Acc_{identify}$ and location metric $F1_{segment}$:
\begin{equation}
Score = \alpha \times Acc_{identify} + (1-\alpha)\times F1_{segment}
\end{equation}
where  $\alpha$ is set to $0.3$ to assign greater weight to the ability to locate the deepfake regions, as it is considered to be the more valuable capability.

The $Acc_{identify}$ measures the model's ability to distinguish between genuine and deepfake audios:
\begin{equation}
    Acc_{identify} =  \frac{TP+TN}{TP+FP+TN+FN} 
\end{equation}
where $TP, FP, TN, FN$ denote the number of test samples detected as true positive, false positive, true negative and false negative, respectively. 

The $F1_{segment}$ evaluates the model's ability to locate the deepfake regions. 
It is the harmonic mean of segment precision and segment recall:
\begin{equation}
\begin{split} 
    P_{segment} &=  \frac{TP_s}{TP_s+FP_s} \\
    R_{segment} &=  \frac{TP_s}{TP_s+FN_s}  \\
    F1_{segment} & = \frac{2}{1/ P_{segment} + 1/R_{segment} }  \\
\end{split} 
\end{equation}
where the true positive is defined as both the reference and model output indicating an event to be active in a segment. 
The $sed\_eval$~\cite{mesaros2016metrics} toolkit is used to calculate the $F1_{segment}$ metric. 
Two temporal resolutions, $1$-Second and $20$-Millisecond, are employed to measure the model's performance at a general resolution and a finer resolution level, respectively.

\label{sec:metric}

\section{Deepfake Detection Model}
\label{sec:model}
We propose a benchmark model for deepfake general audio detection that simultaneously performs deepfake identification and deepfake regions location, as illustrated in Figure~\ref{fig:model}.
A well-performing and efficiently self-supervised EAT~\cite{chen2024eat} is employed to extract frame-level audio representations.

The backbone model is similar to that of Cai et al.~\cite{cai2023dku}, which comprise a ResNet, a two-layer Transformer encoder, a single-layer bidirectional Long Short-Term Memory network (LSTM), and a classification layer. 
Two Convolutional Neural Network (CNN) blocks are positioned respectively before and after the ResNet.
There are 12 blocks in the ResNet, with each block containing two CNN blocks and residual connections. 
The classification layer comprises a fully connected layer, and the output of the classification layer undergoes median filtering to eliminate isolated noisy predictions. 
Subsequently, each frame is predicted as either $1$ or $0$ based on a threshold of $0.5$, where $1$ and $0$ represent genuine and deepfake frames, respectively, for deepfake region location. 
If any frame is predicted as deepfake, the identification result at the clip level is tagged as deepfake.

Furthermore, we explore the impact of multi-task learning by combining frame-level and clip-level fake detection.
An identification layer is added after the classification layer,  dedicated specifically to clip-level deepfake identification.

\begin{table*}[t]
    \renewcommand{\arraystretch}{1.03}
    \centering
    \caption{ 
    Evaluation results of deepfake detection.
    , wherein  ``\textbf{w/o}" and ``\textbf{w}" denotes ``without" and "with", respectively.
    \textbf{``1-Second"} measures the model performance at a general resolution level, while \textbf{``20-Millisecond"} measures the performance of the model at a finer resolution level.
    $\bm{Acc_{identify}}$  evaluates the model's accuracy in identifying genuine / deepfake general audio.
    $\bm{F1_{segment}}$ measures the accuracy of the model in locating deepfake audio regions. 
    $\bm{Score}$ = 0.3 $\times Acc_{identify}$ + 0.7 $\times F1_{segment}$.
    }
    \begin{tabular}{c|c|c|cc|cc}
    \toprule
    \multirow{2}{*}{Acoustic Future}&\multirow{2}{*}{Multi-Task}&\multirow{2}{*}{$Acc_{identify}$} &\multicolumn{2}{c|}{1-Second Resolution}&\multicolumn{2}{c}{20-Millisecond Resolution}  \\
    \cline{4-7}
    & & & $F1_{segment}$ & $Score$& $F1_{segment}$  & $Score$\\
   \bottomrule  

    \multicolumn{7}{c}{Test-Easy}\\
    \midrule
        WavLM & w     &0.710         &0.636         &0.658         &0.616         &0.644         \\
    WavLM &    w/o   &0.790         &0.624         &0.674         &0.580         &0.643         \\
    EAT & w &\textbf{1.000}&\textbf{1.000}&\textbf{1.000}&\textbf{0.980}&\textbf{0.986}\\
    EAT (Proposed)&  w/o &\textbf{1.000}&0.988         &0.992         &\textbf{0.980}&\textbf{0.986}\\
    
    \cline{1-2}
    \multicolumn{2}{c|}{Subjective Evaluation} &0.59  &0.562 &0.571 &0.545 &0.558       \\    
    \bottomrule

    \multicolumn{7}{c}{Test-Hard}\\
    \midrule
    WavLM &w    &0.630         &0.344         &0.430         &0.265         &0.375         \\
     WavLM &   w/o      &0.580         &0.331         &0.406         &0.282         &0.371         \\
    EAT & w  &0.770         &0.738         &0.748         &0.629         &0.671         \\
     EAT (Proposed)&  w/o  &\textbf{0.850}&\textbf{0.834}&\textbf{0.839}&\textbf{0.785}&\textbf{0.805}\\
     \cline{1-2}
     \multicolumn{2}{c|}{Subjective Evaluation}&0.56  &0.368 &0.425 &0.326 &0.396 \\
    \bottomrule

    \multicolumn{7}{c}{Test-Zeroshot}\\
    \midrule   
     WavLM &w     &0.620         &0.283         &0.384         &0.151         &0.292         \\
    WavLM &   w/o   &0.610         &0.255         &0.362         &0.166         &0.299         \\
   EAT &w  &0.700         &0.686         &0.690         &0.644         &0.661         \\
     EAT (Proposed) & w/o  &\textbf{0.720}&\textbf{0.790}&\textbf{0.769}&\textbf{0.782}&\textbf{0.763}\\
    
     \cline{1-2}
   \multicolumn{2}{c|}{Subjective Evaluation} &0.51  &0.293 &0.358 &0.25  &0.328    \\  
    \bottomrule
    \end{tabular}
    \label{tab:result}
\end{table*}

\section{Experiment}
\label{sec:exp}
\subsection{Experiment Setup}

The ResNet, Transformer encoder, and LSTM module share the same hyperparameters as those described by Cai et al.~\cite{cai2023dku}. 
The output dimension of the classification layer is set to $500$, corresponding to $10$-second audio inputs, resulting in a resolution of $20$ ms.

The model is trained for 40 epochs using the AdamW optimizer with learning rate set to $1\times 10^{-4}$.
The entire model, except for the frozen feature extractor EAT, is trained using Binary Cross-Entropy (BCE) loss.
When multi-task learning is employed, the classification for identification is also trained using BCE loss.
The weights for the loss of deepfake region location and identification are set to $0.9$ and $0.1$, respectively.

We utilized the SOTA model in speech deepfake detection, DKU system~\cite{cai2023dku}, as baseline system for comparison, which won the first prize in the ADD2023 competition~\cite{yi2023add}.

\subsection{Subjective Evaluation}
\label{sec:subjective}
To assess the difficulty of the task and validate the necessity of deepfake general audio detection research, we recruited 10 evaluators for human assessment. 
Each evaluator listened to $10$ audio clips from test-easy, test-hard, and test-zeroshot datasets, respectively.
They need to first identify whether the heard audio is a deepfake or not. 
If it is, they are further required to identify the regions they perceive to be deepfake.

The subjective evaluation results are measured using the same metrics as those for the deepfake detection model.
Results are averaged on $10$ evaluators.

\section{ Results}
The experimental results of both machines and subjective evaluation are shown in Table~\ref{tab:result}.
\subsection{Overall performance comparison}
From the results of subjective evaluation, it is evident that deepfake general audio detection presents a highly challenging task for humans. 
Particularly on the Test-Zeroshot set, average accuracy of human binary classification judgments is as low as $0.51$, which is nearly indistinguishable from random guessing.
Hence, the introduction of deepfake general audio detection task is deemed necessary.

It can be observed that the proposed model outperforms the baseline system across all tasks. 
This is attributed to the fact that the baseline system utilizes self-supervised pre-training models trained on speech datasets, extracting acoustical information relevant to speech. 
In contrast, the proposed model employs a pre-training model trained on a large dataset of general audio, capturing features associated with general audio.
This underscores the significant impact of the feature extraction model on deepfake detection task.

In contrast, multi-task learning improved the baseline model but did not enhance the proposed model, suggesting that the proposed model is robust enough and does not require additional training loss design specifically for the identification task.
\subsection{Analysis across 3 test sets}
In detail, the proposed model achieved near-perfect performance on the Test-Easy dataset, with metrics approaching $1$. 
This indicates that the proposed model performs exceptionally well when dealing with test sets that are drawn from the same distribution as the training set.

In the Test-Hard scenario, the duration constraints for reconstructed regions are relaxed, posing greater challenges to the detection model. 
Nevertheless, the proposed model still demonstrates significantly superior performance compared to baseline model and human evaluators, indicating its competitiveness even across test sets with distributions different from the training set. 

However, when evaluated on the hardest Test-Zeroshot dataset, the proposed model exhibits a noticeable decline. 
All metrics $Acc_{identify}$, $F1_{segment}$, and the final $Score$ drop below $0.8$. 
Although the proposed model still outperforms baseline model and subjective evaluation, the performance drop underscores the difficulties the proposed model encounters when facing zero-shot data from different domains. 

Through the analysis of $3$ datasets, it is evident that the proposed model performs better when data is closer in distribution to the training set.
This underscores the current model's \textbf{limitations} in domain adaptation, particularly in zero-shot scenarios, thereby emphasizing domain adaptation as a future research direction for us.

\section{Conclusion}
\label{sec:conclusion}

With the advancement of audio generation tasks, there is an urgent need for deepfake audio detection to prevent potential negative consequences resulting from technological developments.
Therefore, we propose the deepfake general audio detection task, aimed at identifying whether an audio is manipulated or not and locating deepfake segments within it. 
We introduce a manipulation pipeline to automate the acquisition of deepfake general audio, consisting of grounding, masking, and inpainting stages. 
A total of one training set and three test sets are manipulated for training and comprehensively evaluating the deepfake detection model.
Experimental results demonstrate that our proposed model significantly outperforms the state-of-the-art model in speech deepfake detection and subjective evaluation results across all test sets.
However, the current model's constraints lie in domain adaptation.



\section{Acknowledgements}
This work was supported by National Natural Science Foundation of China (Grant No.92048205), the Key Research and Development Program of Jiangsu Province~(No.BE2022059), and Guangxi major science and technology project~(No. AA23062062). 

\bibliographystyle{IEEEtran}
\bibliography{mybib}

\end{document}